\documentclass[aps,preprint,prd,showpacs,nofootinbib]{revtex4-1}
\usepackage{graphicx}
\usepackage[utf8]{inputenc}
\usepackage[T1]{fontenc}
\usepackage{cmap}

\begin{document}

\title{Asymptotic tails of massive gravitons in light of pulsar timing array observations}

\author{R. A. Konoplya}\email{roman.konoplya@gmail.com}
\affiliation{Institute of Physics and Research Centre of Theoretical Physics and Astrophysics, Faculty of Philosophy and Science, Silesian University in Opava, CZ-746 01 Opava, Czech Republic}

\author{A. Zhidenko}\email{olexandr.zhydenko@ufabc.edu.br}
\affiliation{Centro de Matemática, Computação e Cognição (CMCC), Universidade Federal do ABC (UFABC),\\ Rua Abolição, CEP: 09210-180, Santo André, SP, Brazil}

\begin{abstract}
We demonstrate that the late time oscillatory tails of massive gravitons, present in both massive theories of gravity and effectively in extra-dimensional scenarios, could potentially contribute to gravitational waves with very long wavelengths. However, their impact on recent pulsar timing array observations might be relatively small, predominantly consisting of radiation emitted by black holes in our region of the Milky Way.
\end{abstract}

\maketitle

On June 28, 2023 the results of the 15 years observations of millisecond pulsars within our neighbourhood of the Milky Way galaxy were released \cite{NANOGrav:2023gor,NANOGrav:2023hvm,NANOGrav:2023icp,NANOGrav:2023pdq,NANOGrav:2023hfp,Antoniadis:2023lym,Antoniadis:2023xlr,Smarra:2023ljf,Zic:2023gta,Xu:2023wog}.
Gravitational waves of the enormous wavelength (of order of light years) were observed via distortion of the electromagnetic signals from the pulsars.
Although the main candidate for such gigantic gravitational waves are the binary (supermassive) black hole systems in the galactic centers, a number of alternative/additional sources are considered, such as: cosmic inflation or cosmic strings, including the ultra-light dark matter or other kinds of new physical processes.
The binary systems of supermassive black holes have not been directly observed so far and a growing number of works is devoted to further understanding of possible source of the long gravitational waves \cite{Franciolini:2023wjm,Shen:2023pan,Lambiase:2023pxd,Guo:2023hyp,Ellis:2023tsl,Franciolini:2023pbf,Ellis:2023dgf,Ghoshal:2023fhh,Deng:2023btv,Vagnozzi:2023lwo,DiBari:2023upq,Du:2023qvj,Ashoorioon:2022raz,ChoudhuryPandaSami:2023,Choudhury:2023kdb,Huang:2023chx,Cai:2023dls,Inomata:2023zup,Broadhurst:2023tus,Gouttenoire:2023ftk}. However, to the best of our knowledge, one potential source of long gravitational waves was omitted: an effective massive term of a dynamical field under consideration: either gravitational field or matter (e.g. cosmological) field coupled to gravity in some way.
The graviton can acquire an effective massive term owing to the extra-dimensional brane-world scenarios \cite{Seahra:2004fg}, so that not only usual massive theories of gravity would be under consideration.

\begin{figure}
\resizebox{\linewidth}{!}{\includegraphics{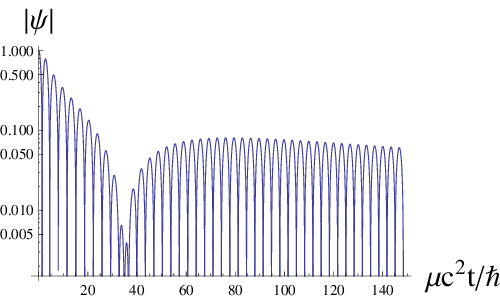}}
\caption{Gravitational wave signal associated with massive modes of $\ell = 2$ axial perturbations of a brane-world black hole \cite{Seahra:2004fg} ($GM\mu/c\hbar=1$): The massive tail dominates at late times and decays slowly.}\label{fig:timedomain}
\end{figure}

\begin{figure}
\resizebox{\linewidth}{!}{\includegraphics{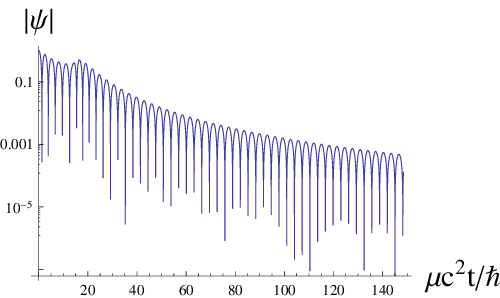}}
\caption{Gravitational wave signal associated with massive modes of $\ell = 2$ axial perturbations of a brane-world black hole \cite{Seahra:2004fg} ($GM\mu/c\hbar=0.1$): The massive tail dominates at late times and decays slowly.}\label{fig:timedomain-smallmass}
\end{figure}

Massive fields in the background of a black hole decay at asymptotically late times according to the universal law, independently on the black hole mass,
\begin{equation}\label{asymptotictail}
\psi\propto t^{-5/6}\sin\left(\frac{\mu c^2}{\hbar} t\right).
\end{equation}
This behavior was observed at late times for the scalar field \cite{KoyamaTomimatsu,Moderski:2001tk}, Proca field \cite{Konoplya:2006gq}, Dirac field \cite{Jing:2004zb}, and massive graviton in Randall-Sundrum-type models \cite{Seahra:2004fg} (see Fig.~\ref{fig:timedomain}).
The power-law enveloping decay rate might be slightly different at the intermediate times, following the ringdown phase. For example, for a massive scalar and gravitational field the decay law at intermediate times is \cite{Hod:1998ra}
\begin{equation}
 \psi\propto t^{-\ell-3/2}\sin\left(\frac{\mu c^2}{\hbar} t\right),
\end{equation}
where $\ell$ is the multipole number. An example of gravitational intermediate tails can be seen in Fig.~\ref{fig:timedomain-smallmass}.

\begin{figure}
\resizebox{\linewidth}{!}{\includegraphics{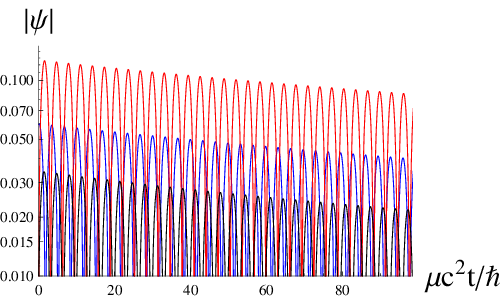}}
\caption{Late-time behavior of the gravitational wave signal associated with massive modes of $\ell = 2$ axial perturbations of brane-world black holes: $GM\mu/c\hbar=0.1$ (red, top) and $GM\mu/c\hbar=1$ (blue, middle) and $\ell = 0$ massive graviton perturbations $GM\mu/c\hbar=1$ (black, bottom). The initial amplitudes are chosen in order to obtain the late-time amplitudes of the same order.}\label{fig:latetime}
\end{figure}

Therefore, we could suppose that the oscillatory law $\sim \sin(\mu c^2 t/\hbar)$ law does not depend on the particular model allowing a graviton to gain an (effective) mass. Evolving the lowest multipole perturbations in the time domain (see Fig.~\ref{fig:latetime}) which we fulfilled here for the massive gravity theory \cite{Brito:2013wya} also confirmed this asymptotic decay law at late times.

It is well-known that the massive fields are short-ranged with the characteristic range
\begin{equation}\label{range}
R\sim\frac{\hbar}{2\mu c}=\frac{\overline{\lambda}_c}{2},
\end{equation}
where $\overline{\lambda}_c$ is the reduced Compton length. However, since gravitons are expected to be ultralight, one could suppose that the range of their massive component interaction could be of the order of light years or more, and can contribute into the Gravitational-Wave Background observed by NANOGrav \cite{NANOGrav:2023gor} and other pulsar timing arrays. This way the extremely large wavelength would be provided simply by a small mass of graviton in the Universe and would not require specific constituents, such as supermassive binary black holes or cosmic strings.

\begin{figure}
\resizebox{\linewidth}{!}{\includegraphics{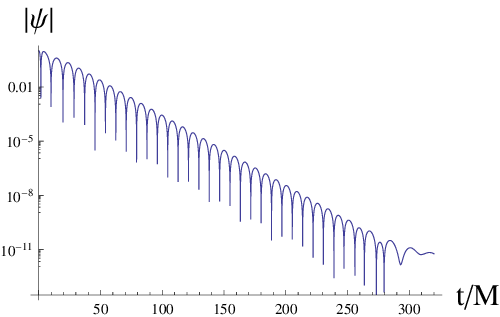}}
\caption{Gravitational wave signal of massless $\ell = 2$ gravitational perturbations of a Schwarzschild black hole ($M$): The signal decays rapidly during the ringdown stage.}\label{fig:ringdown}
\end{figure}

The massless gravitational-wave signal observed by LIGO/VIRGO collaborations \cite{LIGOScientific:2016aoc}, for which the amplitude decreases by many orders of magnitude during the ringdown stage (see Fig.~\ref{fig:ringdown}), is negligibly small comparing to the massive tails. Notice that the time scale of Fig.~\ref{fig:ringdown} defined by the supermassive black hole mass $M$ varies from minutes to several hours while the time scale of Fig.~\ref{fig:timedomain} is of the order of years.

It is crucial for consideration that, unlike the time scale of the ringdown, the late-time decay rate of the massive graviton (\ref{asymptotictail}) is universal and does not depend on the black-hole mass. The similar phenomenon has been observed for wormholes \cite{Churilova:2019qph}. Therefore, other compact objects can contribute to such a signal as well. In Fig.~\ref{fig:latetime} we show the late-time profiles for the perturbations of brane-world black holes \cite{Seahra:2004fg} of different masses together with the monopole massive graviton perturbations \cite{Brito:2013wya}. The amplitude of the oscillations strongly depends on the nature and location of the perturbation source, particular theory of massive gravity, being, thereby, hardly possible to estimate. A comprehensive examination of the perturbation sources in different theories involving massive gravitons lies outside the purview of this letter and is intended for subsequent investigation. Nevertheless, it is clear that despite the initial amplitudes of perturbations of ordinary black holes or other compact objects in our sector of the galaxy being notably smaller than those arising from the black holes formed during galaxy collisions, the extreme proximity of these entities could yield a significant signal amplitude for observers.

Since the asymptotic decay law (\ref{asymptotictail}) is very slow, the Bayesian analysis must be similar to the one for the ultralight dark matter induced signal, which has been performed in \cite{NANOGrav:2023hvm}:
the timing residuals for a pulsar, $h_I$, is written in the form,
\begin{equation}\label{hI}
h_I=\sum_iA_E^i\sin(\omega t+\gamma_E^i)+\sum_iA_P^i\sin(\omega t+\gamma_P^i),
\end{equation}
where $\omega=\mu c^2/\hbar$ and the signal amplitudes, $A_E^i$ and $A_P^i$, and the phases, $\gamma_E^i$ and $\gamma_P^i$ (``E'' and ''P'' subscripts denote Earth and Pulsar term contributions, respectively), must be considered independent, because there is no correlation between independent sources of the gravitational-wave tails.

The energy source we are interested here is the gravitational ringdowns from all the nearby black holes, so that it is not proportional to the dark matter density, like it happens when considering ultralight dark matter as a source. Therefore, the amplitude is not limited anymore by the local abundance of the dark matter, $\rho_{\phi}\approx0.4 GeV/cm^3$, and does not exclude the signal, corresponding the graviton mass of (see Fig.~13 of \cite{NANOGrav:2023hvm})
\begin{equation}\label{estimated}
\mu\sim 2\cdot 10^{-23} eV/c^2 \qquad (\overline{\lambda}_c\sim1~ly).
\end{equation}

It was noted that scalar matter tails, being a source for gravitational perturbations, only slightly wiggle the gravitational tails at late times and do not lead to the oscillating tails \cite{Degollado:2014vsa}. Therefore, if the Gravitational-Wave Background appears due to the ringdown tails, it implies existence of the massive gravitons. Notice that some alternative theories of gravity allow for the gravitational waves with both massless and massive polarizations (e.g., due to non-local curvature corrections \cite{Capozziello:2021bki}). If gravitational perturbations possess massive polarizations along with the massless ones, the massive degrees of freedom will contribute to the large wavelength gravitational oscillations.

It is necessary to compare this estimation with the graviton-mass constraints with the gravitational-wave signals (see~\cite{Piorkowska-Kurpas:2022xmb} for a review).
The LIGO detection of GW150914 provides the upper limit for the graviton mass \cite{LIGOScientific:2016lio},
\begin{equation}\label{GW150914constraint}
\mu<1.2\cdot10^{-22} eV/c^2,
\end{equation}
and the statistical analysis of 24 events shifted the upper limit to \cite{LIGOScientific:2020tif},
\begin{equation}\label{muconstraint}
\mu<1.76\cdot10^{-23} eV/c^2,
\end{equation}
with 90\% credibility, which is of the same order of magnitude as the estimation (\ref{estimated}).

It is worth mentioning that other estimations, based on the temperature and gas density profiles of galaxies, such as Chandra cluster data in X-rays \cite{Gupta:2018pzp}, provide much stronger bounds to the graviton mass, which exclude the graviton masses of (\ref{estimated}). However, starting points of these estimations are the dark matter distributions inferred from a standard cosmological model, which assumes that the graviton mass is zero and, thereby, has no effect on the cosmological evolution. They are also based on the Yukawa gravity model, which does not cover Lorentz invariant massive gravity theories at large distances (see \cite{deRham:2016nuf} for a discussion).

Summarizing, the pulsar timing array observations of the gravitational wave background with the characteristic length of
$$1~ly \lesssim \overline{\lambda}_c \lesssim 10~ly,$$
allow for contribution of massive particles with masses in the range
$$ 2\cdot 10^{-24} eV/c^2 \lesssim \mu \lesssim 2\cdot 10^{-23} eV/c^2.$$
This means that all (non-supermassive) black holes (as well as other compact objects, such as neutron stars) mostly in our corner of the Milky Way could support the long waves observed in \cite{NANOGrav:2023gor,NANOGrav:2023hvm,NANOGrav:2023icp,NANOGrav:2023pdq,NANOGrav:2023hfp,Antoniadis:2023lym,Antoniadis:2023xlr,Smarra:2023ljf,Zic:2023gta,Xu:2023wog} via the mechanism of asymptotic oscillatory tails of massive gravitons described here.
The short-range nature of massive graviton degrees of freedom plays a crucial role in this context. The Compton length of the graviton, ranging from $1$ to $10$ light years, confines its effects to our galaxy. Consequently, the impact of tails from sources in other galaxies, situated at distances of millions of light years, is negligible.
As mentioned earlier, the asymptotic tails is due to the behavior of the wave equation in the far zone, as confirmed by the analysis of the Green functions \cite{KoyamaTomimatsu}, and is not determined by the geometry near the surface of the compact body. Consequently, we must expect the same tails from stars. Although this topic has not been well studied, there some indications that tails must be identical to the black hole case, as, for example, for the Schwarzschild stars considered in \cite{Konoplya:2019nzp}. At the same time, the stellar density near the Sun, which is estimated as $0.004$ stars per cubic light year, is sufficient to provide the gravitational-wave sources.

Despite the dependence of tail amplitudes on the particular model and the initial conditions, from a simple model in Fig.~\ref{fig:timedomain} we see that the tail amplitude can reach several percent of the peak amplitude of the signal.
Thus, within a fundamental or effective model with a massive graviton, the evolution of the gravitational wave will consist of two qualitatively distinct stages: the ringdown stage, where the frequency scales with the mass of the resulting black hole $M$, and the tail stage, where the frequency is determined by the graviton mass.

Taking into account that the amplitude of the signal is inversely proportional to distance, the massive tails within our galaxy can be orders of magnitude larger than the (massless) gravitational-wave amplitudes from events in other galaxies.

If in the future gravitational waves of even longer length will be observed, they could be ascribed to plausible ultra-light massive gravitons with mass smaller than $\mu \approx 2\cdot 10^{-24} eV/c^2$. Since the asymptotic decay law remains consistent for massive particles of various spins, it follows that matter particles should similarly contribute to the signal within the same wavelength. However, a detailed analysis of the sources and each type of particle is necessary to understand the magnitude of their energy content.

It is worth mentioning that, for relatively small astrophysical black holes ($M\sim 6M_{\odot}$),
$$\frac{GM}{c^2}\approx 10^{-12} ly \ll \overline{\lambda}_c \sim 1 ly,$$
the time-domain profile is different than the one shown on Fig.~\ref{fig:timedomain}: The ringdown phase, which lasts fractions of seconds, is negligible at the timescale of years, and the asymptotic tails appear at much later times \cite{KoyamaTomimatsu},
$$t\gg \frac{c^3\overline{\lambda}_c^3}{G^2M^2}.$$
Thus, the intermediate tails dominate at all the times under consideration, i.~e., for
$$t\gg \frac{GM}{c^3}\approx 10^{-12} y.$$

Since the enveloping decay rate, although, usually, slow, might depend on the particular model of massive graviton a thorough consideration of various theories of gravity must be done in order to approach a realistic astrophysical scenario.
Nevertheless, we can expect that the resulting amplitude can be sufficient for observable effects, because there are many sources of the gravitational waves in our galaxy.

\acknowledgments
A.~Z. was supported by Conselho Nacional de Desenvolvimento Científico e Tecnológico (CNPq).

\end{document}